\documentclass[11pt,titlepage]{article}

\usepackage{setspace} 

\usepackage{psfrag,epsfig,amsfonts,amssymb,amsmath,wasysym}

\usepackage[round,numbers,sort&compress]{natbib} 
\usepackage[normalem]{ulem}
\usepackage{color}

\newcommand{\kB}{k}

\title{On the Lubensky-Nelson model of polymer 
translocation through nanopores}

\author{Peter Reimann\thanks{Correspondence: reimann@physik.uni-bielefeld.de}
	\and
	Andreas Meyer
	\and
	Sebastian Getfert\\[0.5cm]
	Universit\"at Bielefeld, \\
	Fakult\"at f\"ur Physik, 33615 Bielefeld, Germany
}

\date{}

\begin{document}

\maketitle

\abstract{
We revisit the one-dimensional stochastic model of 
Lubensky and Nelson [Biophys. J {\bf 77}, 1824 (1999)]
for the electrically driven
translocation of polynucleotides through $\alpha$-hemolysin pores.
We show that the model correctly describes two further important
properties of the experimentally observed translocation time 
distributions, namely their spread (width) and their 
exponential decay.
The resulting overall agreement between theoretical and experimental 
translocation time distributions is thus very good.\\[0.5cm]

}

\clearpage
\section*{Introduction}
\label{s1}
The translocation of biopolymers such as DNA, RNA, or 
polypeptides through protein pores plays a key role 
in various cellular processes \cite{mel03}.
Apart from these biological systems, also artificial,
so-called solid-state nanopores have recently attracted a 
lot of attention due to their promising potential as a 
new generation of fast and cheap DNA sequencing devices 
and other medical diagnostics applications \cite{ven11}.
To achieve such goals, many experimental problems still
have to be solved, and also the theoretical understanding 
and control of those fundamental transport processes 
needs substantial further development.

Here, we reconsider one of the earliest 
and best established theoretical models in this context,
originally introduced in 1999 by Lubensky and 
Nelson \cite{lub99}, and further studied and 
developed in numerous subsequent works, see e.g.
\cite{kaf04,lua05,wan08,che10,li10,lu11,mut10,won10}.
Motivated by the seminal experiments on polynucleotide 
translocation through an $\alpha$-hemolysin pore
by Kasianowicz et al. \cite{kas96}, 
Lubensky and Nelson proposed a theoretical 
description in terms of a one-dimensional 
stochastic model dynamics in a tilted periodic 
potential \cite{lub99}.
While many features of the experimentally
observed translocation time statistics
could indeed be explained remarkably well by their
simple model, the theoretical spread of the translocation 
times underestimated the experimental one by 
about two orders of magnitude \cite{lub99}.
This discrepancy was pointed out once again in 
the review paper \cite{mel03}, but to the best 
of our knowledge has remained a tacitly ignored 
problem of such a model ever since.
To resolve this problem is a first main
issue of our present work.

Since the quantitative details and sometimes
even the qualitative findings notably depend
on the considered pores and polymers, 
we follow Lubensky and Nelson in specifically
focusing on the experimentally best 
studied case of the $\alpha$-hemolysin 
protein pore and polynucleotides  of
single stranded DNA or RNA.
In particular, for this system the translocation
time distributions are quantitatively quite well
documented, not only with respect to their above 
mentioned spread but also with respect to their
decay for large times 
\cite{mel00,mel01,mel02,but07}.
The second main point of our paper
is that the model of Lubensky and Nelson
also correctly reproduces the 
experimentally observed exponential 
decay.

The overall result is a very good comparison
of the complete theoretically predicted
translocation time distributions with
experimentally observed data sets.

\section*{Experimental System}
\label{s2}
\begin{figure}
   \begin{center}
   \epsfxsize=0.8\columnwidth
	\epsfbox{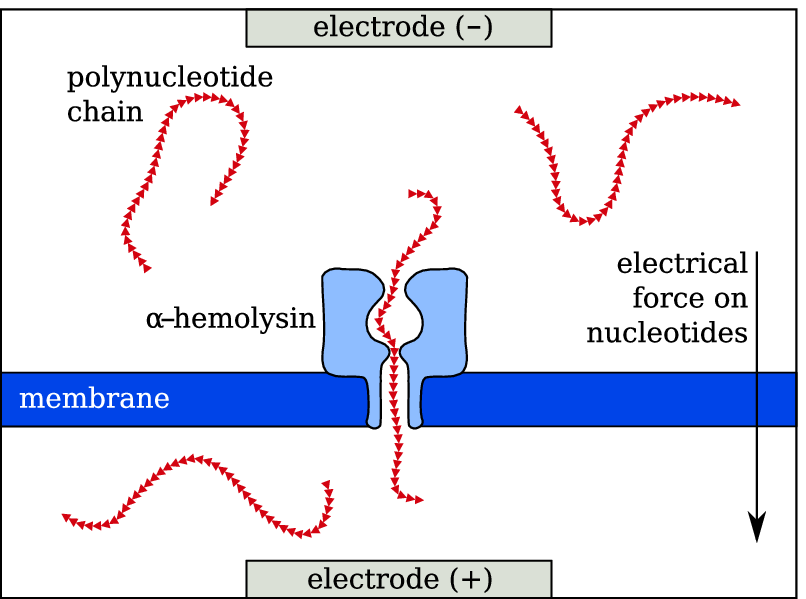} 	
      \caption{}
      \label{fig1}
   \end{center}
\end{figure}

The basic experimental set up is illustrated in Fig. 1.
Charged, single-stranded polynucleotides (DNA or RNA)
in aqueous solution are exposed via electrodes
to an externally applied voltage.
Two fluid compartments are separated by 
a phospholipid membrane 
and are connected by a single 
$\alpha$-hemolysin protein pore.
Since the phospholipid membrane 
is non-conducting, practically the entire voltage drop 
occurs within the pore and its immediate neighborhood.
Whenever a polynucleotide diffusively approaches the
pore from the ``upper'' side in Fig. \ref{fig1}, 
the electrical forces 
direct it into the pore and drive it to the other side of 
the membrane.
Every such translocation process is experimentally observable 
as a reduction of the electrical current through the pore.
Even though the polynucleotides are (practically) 
identical, the durations of the current 
blockades exhibit quite significant statistical 
variations.
The main theoretical task is
to qualitatively explain and quantitatively 
model the experimentally observed
translocation time distributions.
For further details, see, e.g., \cite{mel03,ven11,lub99,kaf04,lua05,wan08,che10,li10,lu11,mut10,%
won10,kas96,mel00,mel01,mel02,but07}.

\section*{Model}
\label{s3}
According to Lubensky and Nelson \cite{lub99}, the polymer
translocation process is modeled by means of a single 
dynamical state variable $x(t)$ (slow/relevant collective coordinate), 
defined as the contour length of that part of the polymer chain 
which already has passed through the pore until time $t$.
In particular, hydrodynamic (dissipative) and steric (entropic)
effects of the chain segments outside the
pore and its immediate neighborhood are considered as negligible.
The most immediate justification of this approximation is that 
otherwise a disagreement with the experimentally observed linear
dependence of the mean translocation time upon the polymer length
\cite{kas96,mel01,mel02,dea02,mel03,dek07,won10}
seems practically unavoidable 
\cite{mil11}\footnote{While this experimental finding 
is beyond any doubt in the case of polynucleotide 
translocation through $\alpha$-hemolysin pores 
\cite{kas96,mel01,mel02,dea02,mel03,dek07,won10}, 
the corresponding results in the case of solid-state 
nanopores \cite{dek07} are contradictory \cite{che04,sto05}.
For this reason the Lubensky-Nelson model may be 
inappropriate in such a case.}.
This general fact is nicely illustrated 
e.g. in Ref. \cite{mut99} 
by means of a model very similar 
in spirit to the one by Lubensky and Nelson, but
in addition taking into account the polymer degrees
of freedom far from the pore region within an 
approximative, accompanying equilibrium description
originally due to \cite{sun96}.

The state variable $x(t)$ is subjected to several
kinds of forces, most notably due to the externally applied 
voltage and the electrostatic, mechanical, and chemical interaction 
of the polymer with the pore walls, but also due to
entropic forces within the pore and its immediate 
neighborhood, generated by the numerous microscopic 
degrees of freedom of the ambient solvent, 
the pore, and the polymer itself.
All those forces can be considered to arise as
minus the derivative of a free-energy type potential 
of mean force $\Phi(x)$.
The remaining effects of the fast molecular degrees of
freedom are approximately modeled as friction (dissipation) and
noise (thermal fluctuations), while inertia effects are
usually negligible on those small lengths and velocity scales.
Altogether, we thus arrive at an overdamped Langevin dynamics
of the well established form \cite{ris84,rei02}
\begin{equation}
\eta \dot x(t) = - \Phi'(x(t)) + \sqrt{2\eta \kB T}\xi(t) \ ,
\label{1}
\end{equation}
where $\xi(t)$ is a delta-correlated Gaussian white noise,
$\eta$ is the friction coefficient, and $\kB T$ the thermal
energy.

In the simplest case of a homopolymer, the force $-\Phi'(x)$ 
remains invariant when the entire polymer is translocated by
the length $a$ of one monomer, i.e. $\Phi'(x+a)=\Phi'(x)$ for all $x$.
As a consequence, $\Phi(x)$ must be a tilted periodic potential,
consisting of a strictly $a$-periodic part $U(x)$ and a
constant ``tilting'' force $F$,
\begin{equation}
\Phi(x)=U(x)-Fx \ .
\label{2}
\end{equation}

Advancing the polymer by one monomer length $a$
changes its (free) energy by $\Phi(x+a)-\Phi(x)=-aF$ 
according to Eq. \ref{2}.
Following Lubensky and Nelson \cite{lub99},
the same change of state is obtained by moving one 
monomer from one to the other end of the polymer 
chain.
The energy required for such a move is $qV$, where
$q$ is the charge of a 
monomer and $V$ the externally applied voltage
(with sign convention as indicated 
in Fig. \ref{fig1}).
We thus can conclude that \cite{lub99}
\begin{equation}
a\,F=-q\, V  \ .
\label{3}
\end{equation}

We remark that the the nominal
charge per nucleotide is equal to
\begin{equation}
q_e= -1.602...\cdot 10^{-19}\,\mbox{C (electron charge).}
\label{3a}
\end{equation}
However, it is by now well established 
\cite{rab05,zha07,gho07,hen00,luo08,bru08}
that due to various electrokinetic effects
of the ambient ionic solution and the pore
(screening, electroosmosis, electrophoresis,
polarization and field confinement mechanisms),
the relevant effective charge $q$ in Eq. \ref{3}
is reduced by roughly a factor of 10 
compared to the nominal (bare) charge 
from Eq. \ref{3a},
i.e.
\begin{equation}
q\approx 0.1\,q_e \ .
\label{3b}
\end{equation}
Due to the above mentioned divers effects which
contribute to the charge renormalization, the 
exact value of $q$ depends, among others, 
on temperature and ion concentrations, 
but also on the specific monomer (nucleotide) 
of which the polynucleotide is composed.

Under the assumption that $V$, $a$, and $q$ are 
(approximately) known,
the force $F$ in Eq. \ref{2} is thus fixed through 
Eq. \ref{3}.
Much more difficult to theoretically estimate from
first principles 
are the friction coefficient $\eta$ in Eq. \ref{1}
and the periodic potential $U(x)$ in Eq. \ref{2}. 
They may thus be considered 
as a model parameter and a model function, respectively,
which remain to be determined by experimental means.

\section*{Velocity and Diffusion}
\label{s4}
As a first quantity of interest we consider the
average translocation velocity $v$ of the polymer
through the pore.
Focusing on not too short polymers, 
``boundary-effects'' while the polymer 
enters and exits the pore are negligible
and $v$ follows as the time- and ensemble-averaged
velocity $\dot x(t)$ from the model in Eq. \ref{1} with an
infinitely extended periodic potential $U(x)$ 
in Eq. \ref{2}.
The analytical solution of this problem goes back 
to Stratonovich \cite{str58} and has subsequently 
been rederived many times, see e.g. chapter 11 
in \cite{ris84}.
Adopting the notation from \cite{rei01,rei02b}, 
this solution takes the form
\begin{equation}
v = \frac{\kB T}{a \eta}
\frac{1-e^{-  aF/\kB T}}{\int_{0}^{a} \frac{dx}{a}\, I (x)}\ ,
\label{4}
\end{equation}
where we have introduced
\begin{eqnarray}
& & I(x) = e^{ \Phi(x)/\kB T}
\int_{x-a}^x \frac{dy}{a}\, e^{- \Phi(y)/\kB T} \ .
\label{5}
\end{eqnarray}

A further quantity of interest is the random spread of the 
translocation velocity (and thus of the translocation time) 
about its mean value $v$, quantified by the diffusion coefficient
\begin{equation}
D = \lim_{t\to\infty}
\frac{\langle[ x(t) - x(0) - vt]^2\rangle}{2t} 
\label{6}
\end{equation}
where $\langle\cdot \rangle$ indicates an 
average over the noise $\xi(t)$ in Eq. \ref{1}
and over the initial positions $x(0)$.

Similarly as for the velocity $v$, the analytical 
result for the diffusion coefficient in a tilted 
periodic potential, Eqs. \ref{1}, \ref{2}, has been 
independently obtained several times.
To the best of our knowledge, the first closed,
exact expression for $D$ is buried in the paper
\cite{par97}.
For the second time, the same problem was 
solved again by Lubensky and Nelson, see 
Appendix B in \cite{lub99}.
Further rediscoveries are due to \cite{lin01}
and \cite{rei01,rei02b}.
While all those results are of course equivalent, 
the actual formulae for $D$ are quite different 
and, with the exception of \cite{rei01,rei02b}, also 
quite involved.
For this reason, the one from \cite{rei01,rei02b} is 
most common, reading
\begin{equation}
D
= \frac{\kB T}{\eta} \,\frac{\int_{0}^{a} \frac{dx}{a}
\, I^2(x)\, J(x)}
{\left[\int_{0}^{a} \frac{dx}{a}\, I(x) \right]^3} \ ,
\label{7}
\end{equation}
where we have introduced
\begin{eqnarray}
& & J(x) = e^{- \Phi(x)/\kB T}
\int_{x}^{x+a} \frac{dy}{a}\, e^{ \Phi(y)/\kB T} \ .
\label{8}
\end{eqnarray}

The main quantity of interest later on will 
be the dimensionless ratio $av/D$ 
(cf. Section ``Spread of Translocation Times'' 
and Ref. \cite{lub99}),
given according to Eqs. \ref{4} and \ref{7} by
\begin{equation}
\frac{av}{D}
= 
\frac{\left[1-e^{-  aF/\kB T}\right]\, 
\left[\int_{0}^{a} \frac{dx}{a}\, I(x) \right]^2} 
{\int_{0}^{a} \frac{dx}{a} \, I^2(x)\, J(x)}
\ .
\label{9}
\end{equation}

A first main result of our paper consists 
in the observation that the leading 
order behavior of Eq. \ref{9}
for small values of $aF/\kB T$ 
takes the form
\begin{equation}
\frac{av}{D} \simeq \frac{aF}{\kB  T} \ ,
\label{10}
\end{equation}
{\em independently of any further details of 
the periodic potential} $U(x)$.
In the simplest case, this result follows by expanding the
left bracket in the numerator of Eq. \ref{9}
to first ($=$leading) order in 
$aF/\kB T$ and evaluating all the remaining 
integrals for $aF/\kB T=0$ 
(leading$=$zeroth order in $aF/\kB T$).
In doing so, the integrals in Eqs. \ref{5} 
and \ref{8} become $x$-independent and, 
as a consequence, the denominator in Eq. \ref{9} 
becomes equal to the right bracket in the 
numerator. 
An analogous (but more tedious) expansion
resulting in \ref{10} is also contained 
in \cite{lub99}.
Both expansions, however, become questionable 
in the weak noise limit (small thermal 
energy $\kB T$), since the expansion 
coefficients in general will inherit
exponentially large values from the integrands
$\exp\{\pm [U(x)-U(y)]/\kB T\}$ contributing
via Eqs. \ref{2}, \ref{5}, \ref{6} to the 
multiple integrals in \ref{9}.

Our first remark is that Eq. \ref{10} in fact 
still remains valid
for asymptotically weak noise ($\kB T\to 0$)
and not too large 
$F$-values, 
so that
the dynamics in Eqs. \ref{1}, \ref{2}
is governed by rare, thermally activated 
transitions between metastable states:
In this case, the dynamics can be approximately
described by a one-dimensional random walk between discrete
sites at distance $a$ with certain forward and backward
hopping rates $r_+$ and $r_{-}$, respectively.
As a consequence \cite{rei02}, one obtains 
$v=a(r_+-r_-)$ and $D=a^2 (r_++r_-)/2$.
Furthermore, detailed balance symmetry \cite{rei02}
implies for the forward and backward rates the relation
$r_+/r_{-}=\exp\{aF/\kB T\}$.
We thus obtain the asymptotically exact
result
\begin{equation}
\frac{av}{D} = 2\,
\frac{1-e^{-aF/\kB T}}{1+e^{-aF/\kB T}}=2\,\tanh(aF/2\kB T) \ ,
\label{9b}
\end{equation}
independently of any further 
details of $U(x)$.
In particular, for small $aF/\kB T$ one readily 
recovers Eq. \ref{10}.

Our second remark is that both for asymptotically large
$F$ and for asymptotically large $\kB T$, the effects
of the periodic potential $U(x)$ in Eq. \ref{2} become
negligible \cite{rei01,rei02b} with the consequence that
the exact expression in Eq. \ref{9} approaches once 
again the asymptotics of Eq. \ref{10}.
The same conclusion is of course recovered if the 
variations of the periodic potential $U(x)$ itself 
become negligibly small.

Our last remark is that the diffusion coefficient $D$ 
as a function of the tilt $F$ develops an arbitrarily 
pronounced maximum for sufficiently small $\kB T$, 
see \cite{rei01,rei02b}.
Nevertheless, closer inspection along the lines of
\cite{rei01,rei02b} shows that the  ratio $D/av$ remains a 
strictly decreasing function of $F$ within the
neighborhood of the maximum of $D$.

\section*{Translocation time distribution and exponential decay}
\label{s5}
A ``successful translocation event'' starts when 
the polymer enters the pore from one side 
and ends when it exits at the other side.
In contrast, if the polymer exits at the same 
side as it entered, we are dealing with an 
``unsuccessful translocation attempt''.
Following Lubensky and Nelson, we ignore 
unsuccessful attempts and henceforth only 
consider the successful events.
Regarding the experimental identification of 
such events see e.g. \cite{but07}.
The statistical distribution of their duration
is the quantity of foremost interest from now 
on.

A main achievement of Lubensky and Nelson's work \cite{lub99}
is an analytical approximation for the distribution
(probability density) $\psi(t)$ of translocation times.
In fact, the approximation becomes
asymptotically exact for sufficiently large 
numbers $N$ of monomers, i.e.
\begin{equation}
N=L/a \gg 1 \ ,
\label{11}
\end{equation}
where $a$ and $L$ denote the lengths of one monomer and
of the entire polymer, respectively.

Remarkably enough, but also quite plausible at second 
glance, the only parameters entering the translocation 
time distribution $\psi(t)$ are the polymer length $L$
and the velocity $v$ and diffusion coefficient 
$D$ of the corresponding, infinitely extended dynamics.
Furthermore, it is convenient to employ the rescaled,
dimensionless time 
\begin{equation}
\tau = \frac{v\, t}{L}
\label{12}
\end{equation}
and the dimensionless auxiliary parameter
\begin{equation}
\kappa  =\frac{4\, D}{v\, L} \ .
\label{13}
\end{equation}
Referring to Appendix A of \cite{lub99} for the detailed
calculations, the final analytical expression 
provided by eq. (A6) in \cite{lub99} and can be 
rewritten 
in the form
\begin{equation}
\psi(t) = \frac{c}{\tau^{3/2}}
\sum_{n=1,3,5,...}
\frac{\frac{n^2}{\kappa \,\tau}-\frac{1}{2}}
{\exp\left\{\frac{2(n-1)}{\kappa }
+\frac{(\tau-n)^2}{\kappa  \tau}\right\}}
\label{14}
\end{equation}
where the normalization constan $c$ is given 
by\footnote{There is a typo on the right hand side 
of eq. (A6) in \cite{lub99}: the first factor 
$2$ should be replaced by $1/2$} 
\begin{equation}
c =\frac{v}{L}\, \sqrt{\frac{\kappa}{\pi}}\, 
[ 1-e^{-4/\kappa} ] \ .
\label{14a}
\end{equation}
In other words, the translocation time distribution 
$\psi(t)$ actually does not depend separately on all 
three parameters $L$, $v$, and $D$, 
but only on the two specific combinations 
$v/L$ and $\kappa=4D/vL$.

The behavior of Eq. \ref{14} for large $t$
is not obvious at all. In particular, for large $\tau$
the summands on the right hand side are negative up to quite large $n$-values,
while those for even larger $n$ are positive.
In view of the property $\psi(t)\geq 0$ and the expected 
asymptotics $\psi(t)\to 0$
for $t\to\infty$, there must be a very fragile
cancellation of positive and negative summands.
On the other hand, we observe that, according to
eq. (A2) in \cite{lub99}, an asymptotically exponential 
decay of the right hand side in Eq. \ref{14} may 
be conjectured with a decay rate 
\begin{equation}
\lambda = (\pi/2)^2 \kappa  + 1/\kappa  \ .
\label{15}
\end{equation}
In view of this surmise, we thus may rewrite Eq. \ref{14}
in the following equivalent form
\begin{equation}
\psi(t) = c' \ 
g(\kappa \, \tau) 
\ e^{-\lambda \tau} \ ,
\label{16}
\end{equation}
where $c'$ is another normalization constant,
and where the auxiliary function $g(x)$ 
is defined as follows
\begin{eqnarray}
g(x) & = & 
0.45731\, \frac{e^{x\pi^2 /4}}{x^{3/2}}
\sum_{n=1,3,5,...}
\frac{\frac{n^2}{x}-\frac{1}{2}}{e^{n^2/x}}  \ .
\label{17}
\end{eqnarray}

As can be seen from Fig. \ref{fig2}, the function $g(x)$ 
converges towards a finite limit for 
$x\to\infty$ and the specific numerical 
factor $0.45731$ in Eq. \ref{17} has been chosen 
so that  this limit is (practically) unity.

This brings us to the next main result 
of our paper:
According to Eq. \ref{16} and Fig. \ref{fig2}, 
the distribution of
translocation times $\psi(t)$ predicted
by the model of Lubensky and Nelson
exhibits an  exponential decay for large times, 
in agreement with the experimental findings from
\cite{mel00,mel01,mel02,but07}.

We finally remark that equation \ref{16} 
together with Fig. \ref{fig2} 
and Eqs. \ref{12}, \ref{13}, \ref{15}
provide a quite detailed qualitative
picture of the translocation time 
distribution $\psi(t)$, 
and how it depends on the parameters 
$v/L$ and $\kappa$:
Initially, $\psi(t)$ remains close to zero,
then increases quite steeply up to a maximum 
at $t_{max}=\tau_{max}L/v$, where $\tau_{max}$ solves 
$\kappa \, g'(\kappa \tau_{max})=\lambda\, g(\kappa \tau_{max})$,
and finally approaches an exponential decay.

Numerically, for any given set of parameters $v/L$ 
and $\kappa$ the quantitative evaluation of $\psi(t)$
according to Eqs. \ref{12}-\ref{14a} is straightforward.
In particular, for $\kappa\to 0$ (vanishing thermal 
fluctuations) one recovers the expected deterministic 
limit $\psi(t)\to\delta(t-L/v)$.
Typical examples will be presented in Section
``Comparison with Experiments'' below.

\section*{Spread of translocation times}
\label{s6}
Concerning a quantitative comparison between
the theoretical prediction from Eq. \ref{14} 
and experimental data, we first 
observe that there are two fit parameters:
One (namely $L/v$ in Eq. \ref{12})
amounts to a quite trivial rescaling
of time, which can be readily
fixed e.g. by fitting the theoretical peak
of $\psi(t)$ to the experimentally 
observed one.
The remaining, second fit parameter is $\kappa $ 
from Eq. \ref{13}, which can be determined
by the following very convenient procedure 
originally due to Lubensky and Nelson 
\cite{lub99}:
In a first step, the peak position $t_{max}$ of 
the experimentally observed $\psi(t)$ is determined, 
formally defined via $\psi'(t_{max})=0$.
Then the two times $t_L$ and $t_R$ 
to the left (L) and to the right (R) 
of $t_{max}$ are determined according to 
\begin{equation}
\psi(t_{L,R})=e^{-1/2}\, \psi(t_{max}) \simeq 0.606 \cdot  \psi(t_{max}) \ .
\label{18}
\end{equation}
In other words, $t_R-t_L$ quantifies
the width of the experimentally observed 
peak of $\psi(t)$.
Observing that the ratio 
$(t_R-t_L)/t_{max}$
is independent of the chosen 
units of time, we can readily
recast the approximative relation
from Fig. 3 of Lubensky and Nelson's
work \cite{lub99} into the form
$(t_R-t_L)/t_{max}\simeq \sqrt{2\kappa }$.
In fact, this approximation becomes exact for
asymptotically small $(t_R-t_L)/t_{max}$
and remains quite accurate as long as 
$(t_R-t_L)/t_{max}\leq 1$.
In other words, we obtain the quite accurate estimate
\begin{equation}
\kappa \simeq \frac{1}{2}\left(\frac{t_R-t_L}{t_{max}}\right)^2 
\ \ \mbox{provided} \ \ \frac{t_R-t_L}{t_{max}}\leq 1 \ .
\label{19}
\end{equation}

On the other hand, combining Eqs. 
\ref{3}, \ref{10}, \ref{11}, and \ref{13}
yields the relation
\begin{equation}
\frac{\kB T}{(-q)\, V} \simeq \frac{\kappa\, N}{4}\ \mbox{provided} \ \frac{\kappa\, N}{4}
\geq 1 \ .
\label{20}
\end{equation}

{From} the experimental data in Fig. 2 of 
Kasianowicz et al. \cite{kas96} 
one readily reads off $(t_R-t_L)/t_{max}\simeq 1$ \cite{lub99},
implying with Eq. \ref{19} that $\kappa \simeq 0.5$.
Taking into account that $N=210$ in the experiments 
from \cite{kas96}, the right hand side of Eq. \ref{20} 
amounts to $\kappa N/4\simeq 26$.
Hence the condition $\kappa N/4\geq 1$ is fulfilled
and we can apply Eq. \ref{20} to conclude that 
$\kB T/qV\simeq 26$.
On the other hand, using $T\simeq 293\,$K 
(room temperature), 
$V=120\,$mV (experimental voltage from \cite{kas96}), 
and the nominal charge 
$q=q_e\simeq - 1.6\cdot 10^{-19}\,C$ 
per nucleotide from \cite{lub99},
we obtain the result $\kB T/(-q)V\simeq 0.21$.
In view of the discprepancy with the relation
$\kB T/(-q)V\simeq 26$ following from Eq. \ref{20}, 
Lubensky and Nelson concluded that their model, Eq.
\ref{1}, which implied Eq. \ref{20}, 
was inconsistent with the experimental facts.

In the following, we argue that this conclusion is
not tenable. Rather, the prediction from Eq. \ref{20} of
the model, Eq. \ref{1}, agrees quite well with the 
experimental findings.
To this end, we first evaluate the right hand side
of Eq. \ref{20} by means of the results from several more
recent, and therefore possibly more accurate
experiments than in the original work of Kasianowicz et al.
\cite{kas96}: From the two data sets of Meller at al.
displayed in Fig. 2 of their work \cite{mel00} 
(see also Fig. 3 in \cite{mel01} and Fig. 6 
in \cite{mel02}) one can infer with Eq. \ref{19} that
$\kappa\simeq 0.09$ and $\kappa\simeq 0.11$, respectively
(see also next Section).
With $N=100$ \cite{mel00} it follows that $\kappa N/4\simeq 2.3$ 
and $\kappa N/4\simeq 2.8$, respectively.
Likewise, one can infer from the data
in Fig. 2 of Bates et al. \cite{bat03}
that $\kappa\simeq 0.17$ 
and hence with $N=60$ that
$\kappa N/4\simeq 2.6$.
Finally the data from Fig. 5 of Butler et al. 
\cite{but07} imply that $\kappa\simeq 0.19$ and 
with $N=50$ that $\kappa N/4\simeq 2.4$.

We remark that in all those experiments homopolymers have
been used and that we do not know of any further 
data of this kind 
(histograms of translocation durations for homopolymers)
in the literature.
For the sake of completeness, we may also include 
here the data for heteropolymers with $N=92$ 
from Fig. 2c in Maglia et al.
\cite{mag08}, yielding $\kappa=0.13$ and hence 
$\kappa N/4\simeq 3.0$.

Turning to the quantity $\kB T/(-q)V$ appearing
on the left hand side of Eq. \ref{20},
we observe that the voltage $V=120\,$mV, and the
temperature $T\simeq 293\,$K was (approximately)
the same in all the experiments from 
\cite{kas96,mel00,mel01,mel02,bat03,but07,mag08}.
Using the nominal charge $q=q_e$ per nucleotide from
\cite{lub99}, one thus recovers the same result 
$\kB T/(-q)V\simeq 0.21$ as before, and hence 
Eq. \ref{20} still seems to be violated.
However, according to 
\cite{rab05,zha07,gho07,hen00,luo08,bru08}
a much more realistic estimate follows
by employing the appropriately
renormalized effective charge from 
Eq. \ref{3b}, namely $\kB T/(-q)V\simeq 2.1$.
As a consequence, the theoretically predicted 
relation in Eq. \ref{20} is satisfied quite 
well by all the more recent pertinent experiments 
\cite{mel00,mel01,mel02,bat03,but07,mag08}.

Taking the model of Lubensky and Nelson and thus 
Eq. \ref{20} for granted, the above
findings for $\kappa N/4$ may now 
in turn be used to estimate the renormalized 
charge $q$ more accurately than in Eq. \ref{3b}.
Since temperature, voltage, buffer etc. 
were almost the same in all cases,
the resulting differences in $q$ must be
mainly due to the different nucleotides
(see also below Eq. \ref{3b}).

With respect to the earlier experiment by Kasianowicz et al.
\cite{kas96}, a further reduction of the effective
charge by another factor of 10 would be a possible,
though not very satisfying, explanation (see also 
next Section).
A more likely reason seems to be connected
with the considerably larger $\kappa N/4$-value
compared to the more recent experiments.
In other words, the experimentally observed spread
of the translocation times is unusually large.
Indeed, one generally expects that the experimentally observed 
spreads of the translocation times still somewhat
overestimate the purely diffusive effects accounted 
for in the theory. 
One possible reason of why the observed spread was
particularly large in \cite{kas96} may be that
the data analysis and pre-processing according to
their detailed current blockade signatures
was not yet as sophisticated as e.g. in 
\cite{mel00,mel01,mel02,but07}.
Another possible factor is the smaller measurement bandwidth
of about $24\,$KHz for the experiments \cite{kas96},
compared to $100\,$KHz for \cite{mel00,mel01,mel02,bat03} 
and $50\,$KHz for \cite{but07}.
As will be argued in the next section, the most
plausible explanation is a relatively large spread 
of the polynucleotide lengths of the samples used
by Kasianowicz et al. \cite{kas96}.

\section*{Comparison with experiments}
\label{s7}
The purpose of this Section is a 
comparison between the complete 
translocation time distributions
for several of the experiments already considered 
in the previous Section and the corresponding
theoretical distributions.
Quite surprisingly, such a detailed
quantitative comparison does not seem to exist 
in the previous literature known to us.

As a first example, Fig. 3 presents the experimental 
data reported by Butler et al. \cite{but07} for single-stranded 
RNA $rC_{50}$ polynucleotides (i.e. $N=50$ in Eq. \ref{11}).
The theoretical translocation time distribution has been
obtained as described in the previous Section:
First, one readily reads off from the experimental 
data points that $(t_R-t_L)/t_{max}\simeq 0.61$, 
yielding with Eq. \ref{19} the estimate $\kappa\simeq 0.19$.
Then, the time-scale parameter $L/v$ in Eq. \ref{12}
is adapted so as to optimally fit the peak position 
of the experimental data,
resulting in the estimate $L/v\simeq 0.2\,$ms.
According to Fig. \ref{fig3}, the theoretical 
$\psi(t)$ obtained in this way agrees very well 
with the experimental findings with the exception 
of the times $t$ smaller than about $0.1\,$ms.
In fact, the corresponding experimental data were denoted 
in Ref. \cite{but07} as ``ambiguous signals'', possibly 
caused by ``retraction of the threaded configuration 
back into the vestibule configuration, very rapid
translocation, or translocation of short 
polynucleotide fragments.
Due to the ambiguity in the interpretation of 
these short Deep states, we only designated Deep 
states with durations longer than the minimum 
as translocations.''
This ambiguity in the interpretation
of the experimental current blockades
(``Deep states'')
seems to us a sufficiently convincing 
explanation of the deviations from the 
theoretical curve at small times $t$.
An apparently rather similar situation 
for small $t$ has in fact been discussed 
already in \cite{kas96,lub99,but06}.
In view of this ambiguity
for $t<0.1\,$ms we fitted in Fig. \ref{fig3}
the scaling factor $c$ of the theoretical curve 
\ref{14} as well as possible to the 
experimental data for $t>0.1\,$ms rather 
than normalizing it according to \ref{14a}.

As a second example, Fig. \ref{fig4} shows the
data of Bates et al. \cite{bat03} for single-stranded 
DNA $dA_{60}$ polynucleotides (i.e. $N=60$ in Eq. \ref{11}).
Theoretically, we proceeded as before with
$(t_R-t_L)/t_{max}\simeq 0.58$, 
$\kappa\simeq 0.17$, and $L/v\simeq 0.42\,$ms.
Again, the overall agreement between theory and 
experiment is very nice with the exception 
of large times 
$t$\footnote{In fact, the authors of Ref. \cite{bat03} 
mention that the tail of the experimentally observed 
distribution extends to even
much longer times without providing the actual data.}.

As pointed out by an anonymous referee, the latter
disagreement 
can be naturally explained by the well-known directionality of 
the polynucleotide's sugar-phosphate backbone, 
resulting in two different 
translocation time distributions depending on whether
the DNA enters the pore with its 3' or 5' end first
\cite{kas96,lub99,ake99,wan04,mat05,but06,but07,pur08,muz10,aks10}.
Denoting the probabilities of 3' and 5' entries by $p$ and 
$1-p$ and the concomitant two distributions 
by $\psi_1(t)$ and $\psi_2(t)$,
the total (experimentally observed) distributions 
is given by
\begin{equation}
\psi(t)=p\psi_1(t)+(1-p)\psi_2(t) \ .
\label{16b}
\end{equation}
Within the model of Lubensky and Nelson, 
$\psi_{1}(t)$ and $\psi_2(t)$ are both of the 
form \ref{14}. 
For both of them, the parameter $\kappa$ must
be the same according to
Eqs. \ref{3}, \ref{10}, \ref{13}
under the plausible assumption that
the effective charge $q$ is (approximately) 
the same for both DNA orientations
(see also below Eq. \ref{3b}).
On the other hand,
there may in general be two different 
time scales $L/v_1$ and $L/v_2$, and 
likewise for $\tau$ in Eq. \ref{12}.
Along these lines, the best fit to the
experimental data of Bates et al. \cite{bat03}
was obtained for $\kappa\simeq 0.18$,
$L/v_1\simeq 0.42\,$ms, 
$L/v_2\simeq 0.82\,$ms, and $p\simeq 0.81$.
According to Fig. \ref{fig5}, the resulting
agreement between experiment and theory is
indeed very good.
We remark that while quantitative experimental 
estimates for the two ``extra parameters''
$p$ and $v_1/v_2$ of the extended theory \ref{16b}
do not seem available, the ``event diagrams'' 
for $dA_{50}$ in Fig. 3 of the paper 
\cite{but07} 
and for $dA_{100}$ in Fig. 2 of the paper 
\cite{wan04} 
qualitatively compare very 
favorably with our above findings $p\simeq 0.81$ and 
$v_1/v_2\simeq 2$ for $dA_{60}$.
We finally remark that also in the works 
\cite{mel00,wan04} two ``groups'' of $dA_{100}$ 
translocation events were identified,
``group 1'' containing about 80\% of the
events (at $20^\circ\,$C) \cite{mel00},
and that the two ``groups'' can be attributed 
to the two different DNA orientations 
\cite{wan04,but06}.

As expected, the already very good agreement 
in Fig. \ref{fig3} with the  $rC_{50}$ data 
by Butler et al. \cite{but07} could not be 
notably improved any more by means of the
extended theory from Eq. \ref{16b}.
This is consistent with the ``event diagram''
for $rC_{50}$ in Fig. 3 of the paper 
\cite{but07}, evidencing that while the
RNA may indeed again exhibit two 
different orientations, the translocation
time distributions happen to be very 
similar for both of them.

We finally return to the experiment of 
Kasianowicz et al. \cite{kas96}, using
poly[U] samples with a nominal length
of $N=210$ nucleotides.
In a later work \cite{ake99}, the same group
used a poly[U] sample whose length distribution
was specified as $N=150\pm 50$ nucleotides.
According to a personal communication by
one of the authors (D. Branton, Harvard University),
a comparable polydispersity of $N=210\pm 70$
may thus be considered as 
quite plausible also in the earlier work \cite{kas96}.
Theoretically, we took into account this fact by
including on the right hand side of Eq. \ref{16b}
an integral over a Gaussian length distribution.
Formally, this is achieved by replacing
$L$ in the definitions \ref{12} and \ref{13}
by $z\,L$,
where $z>0$ is Gaussian 
distributed\footnote{Since $z\leq 0$ is ruled out, 
the Gaussian must be truncated and properly 
renormalized.}
with average $1$ and standard deviation $1/3$,
and likewise for the two time scales $L/v_1$,
$L/v_2$ entering into Eq. \ref{16b}.
Fig. \ref{fig6} shows our best fit to the
experimental data of Kasianowicz et al. \cite{kas96},
obtained for $\kappa\simeq 0.1$,
$L/v_1\simeq 0.31\,$ms, 
$L/v_2\simeq 1.3\,$ms, and $p\simeq 0.58$.
The agreement is obviously very good,
except for the leftmost data point in Fig. \ref{fig6}.
In fact, there are additional experimental data 
points at even smaller $t$-values with
$\psi$-values beyond the range displayed in 
Fig. \ref{fig6}.
Similarly as in Fig. \ref{fig3}, 
they are commonly considered to be due to 
polymers that partially entered the channel 
but then retracted rather than actually traversed
the channel \cite{kas96,lub99,but06,but07}.
Such events are not covered by our 
present theory, explaining the disagreement
in Fig. \ref{fig6} at small $t$.
Accordingly, the experimental data in Fig. \ref{fig6} 
have not been normalized to unity but rather so
that the agreement with the theory was optimal.

We remark that such ``unsuccessful 
translocation attempts'' could be sorted out 
in the experimental data in Figs. \ref{fig4} 
and \ref{fig5} thanks to their very specific current 
blockade signatures \cite{bat03}.
Hence the theory explains the data even at
small $t$.
A similar identification of such events
in the experiments by Kasianowicz 
et al. \cite{kas96} and by Butler 
et al. \cite{but07} was apparently 
not possible.

Returning to Fig. \ref{fig6}, the obtained
fit $\kappa=0.1$ implies that $\kappa N/4=5.25$
and thus with Eqs. \ref{20} and \ref{3a} that
$q\approx 0.04\, q_e$.
Such an effective charge is smaller than the 
estimate from equation \ref{3b} but still reasonably close
to our previously obtained 
$q$-values\footnote{Note that also the
``blocking currents'' may vary substantially
for different nucleotides and that this may 
be closely related to variations of $q$
\cite{zha07}.}.
We also remark that by assuming a larger
polydispersity of $N=210\pm 95$, the
data from Fig. \ref{fig6} could be fitted 
practically equally well but now with
$\kappa=0.05$ and hence $q\approx 0.08\, q_e$,
and likewise for $N=210\pm 120$ and any
$q > 0.5\, q_e$.
In contrast, without any polydispersity 
the best fit becomes somewhat worse than in 
Fig. \ref{fig6} and $q\approx 0.015\, q_e$ 
becomes unrealistically small.

Finally, we also fitted the above extended 
model to the experimental data sets from 
Figs. \ref{fig3} and \ref{fig5} but for
any non-negligible polydispersity 
this always resulted in a less good 
agreement than without any polydispersity.

\section*{Discussion}
\label{s8}
In their seminal work \cite{lub99}, Lubensky and Nelson
showed that their simple one-dimensional stochastic model, Eqs.
\ref{1}-\ref{3}, captures many of the
experimental observations on polynucleotide
translocation through $\alpha$-hemolysin 
nanopores.
However, the quantitative behavior of the
experimental translocation time distribution 
of Kasianowicz et al. \cite{kas96} could not be satisfactorily 
explained.
Here, we resolved this long standing problem by taking into
account that due to various electrokinetic effects 
the relevant effective charge of the nucleotides
is substantially reduced compared to their
bare (nominal) charge.

A second main point of our work was to show that
the model of Lubensky and Nelson
implies an asymptotically exponential decay 
of the translocation time distribution, 
in agreement with the experimental results 
from Refs. \cite{mel00,mel01,mel02,but07}.

Finally, we compared the complete
theoretically predicted translocation time 
distributions with experimental 
distributions from the literature.
Taking into account the directionality
of polynucleotides, the non-negligible
polydispersity in the experiment 
by Kasianowicz et al. \cite{kas96}, 
and the fact that ``unsuccessful translocation 
attempts'' are not covered by the theory,
the model of Lubensky and Nelson
explains the experimental observations
remarkably well.

Regarding alternative, more sophisticated descriptions
e.g. in term of bead-spring models with many degrees of
freedom \cite{mil11}, one naturally expects that --
at least within certain regimes or limits
of the various model parameters -- 
the main features of the one-dimensional model of 
Lubensky and Nelson should be recovered.
Particularly important such features are the
experimentally observed linear dependence of the mean 
translocation time upon the polymer length
(see also Section ``Model'') and the main quantitative
characteristics of the observed translocation time 
distributions, most notably their spread and 
exponential decay.
The essential open question with respect to those
more sophisticated models is then, whether
the numerous extra degrees of freedom in order
to describe the hydrodynamic and entropic effects outside the
immediate pore region still play a significant 
role within the above mentioned, relevant model 
parameter regimes.

\begin{center}
\vspace{-5mm}
---------------------------
\vspace{-4mm}
\end{center}
We thank D. Branton for a very helpful e-mail exchange
regarding Ref. \cite{kas96}.
This work was supported by Deutsche Forschungsgemeinschaft 
under SFB 613 and RE1344/8-1.

\clearpage
\section*{Figure Legends}
\subsubsection*{Figure~\ref{fig1}}
Schematic illustration of the experimental set up:
polynucleotide chains (single-stranded DNA or RNA) 
translocate through 
an $\alpha$-hemolysin protein pore due to the externally 
imposed voltage difference between the electrodes.

\subsubsection*{Figure~\ref{fig2}}
The function $g(x)$ obtained by numerically 
evaluating Eq. \ref{17}.

\subsubsection*{Figure~\ref{fig3}}
Line: Theoretical translocation time distribution 
$\psi(t)$ according to Eqs. \ref{12}-\ref{14a}
with $\kappa=0.19$ and $L/v=0.2\,$ms.
Symbols: Experimentally observed translocation times,
adopted from Fig. 5 of Butler et al. \cite{but07}.

\subsubsection*{Figure~\ref{fig4}}
Line: Theoretical translocation time distribution 
$\psi(t)$ according to Eqs. \ref{12}-\ref{14a}
with $\kappa=0.17$ and $L/v=0.42\,$ms.
Symbols: Experimentally observed translocation times,
adopted from Fig. 2 of Bates et al. \cite{bat03}.

\subsubsection*{Figure~\ref{fig5}}
Line: Theoretical translocation time distribution 
$\psi(t)$ according to Eqs. \ref{16b}
(see also main text for details)
with $\kappa=0.18$,
$L/v_1= 0.42\,$ms, 
$L/v_2= 0.82\,$ms, and $p=0.81$.
Symbols: Experimentally observed translocation times,
adopted from Fig. 2 of Bates et al. \cite{bat03}.

\subsubsection*{Figure~\ref{fig6}}
Line: Theoretical translocation time distribution 
$\psi(t)$, accounting for polydispersity 
by randomizing Eqs. \ref{16b}
(see main text)
with $\kappa= 0.1$,
$L/v_1= 0.31\,$ms, 
$L/v_2= 1.3\,$ms, $p=0.58$.
Symbols: Experimentally observed translocation times,
adopted from Fig. 2 of Kasianowicz et al. \cite{kas96}.

\clearpage

\clearpage
\begin{figure}
   \begin{center}
   \epsfxsize=0.8\columnwidth
	\epsfbox{fig2.eps} 	
      \caption{}
      \label{fig2}
   \end{center}
\end{figure}

\clearpage
\begin{figure}
   \begin{center}
   \epsfxsize=0.8\columnwidth
	\epsfbox{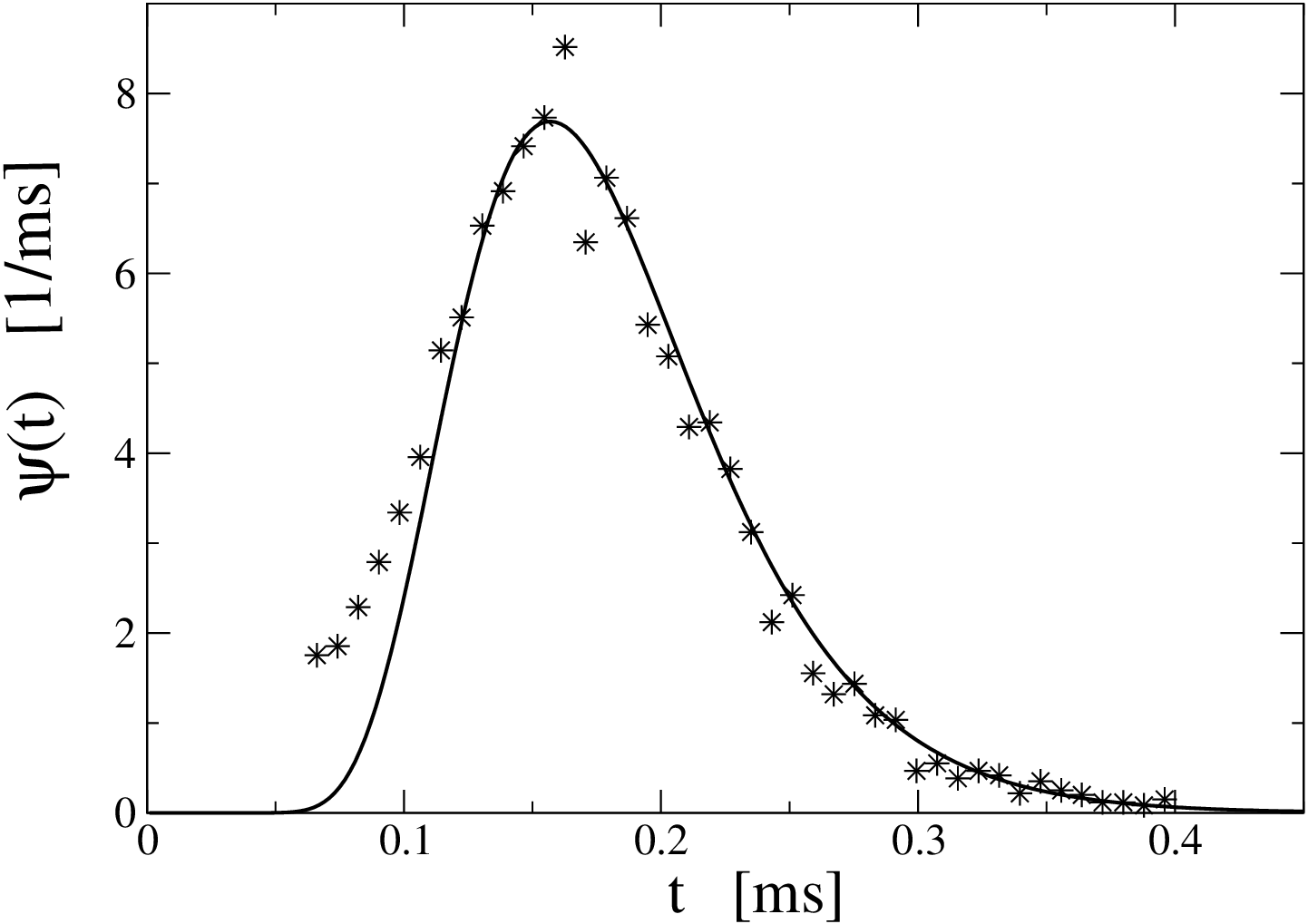} 	
      \caption{}
      \label{fig3}
   \end{center}
\end{figure}

\clearpage
\begin{figure}
   \begin{center}
   \epsfxsize=0.8\columnwidth
	\epsfbox{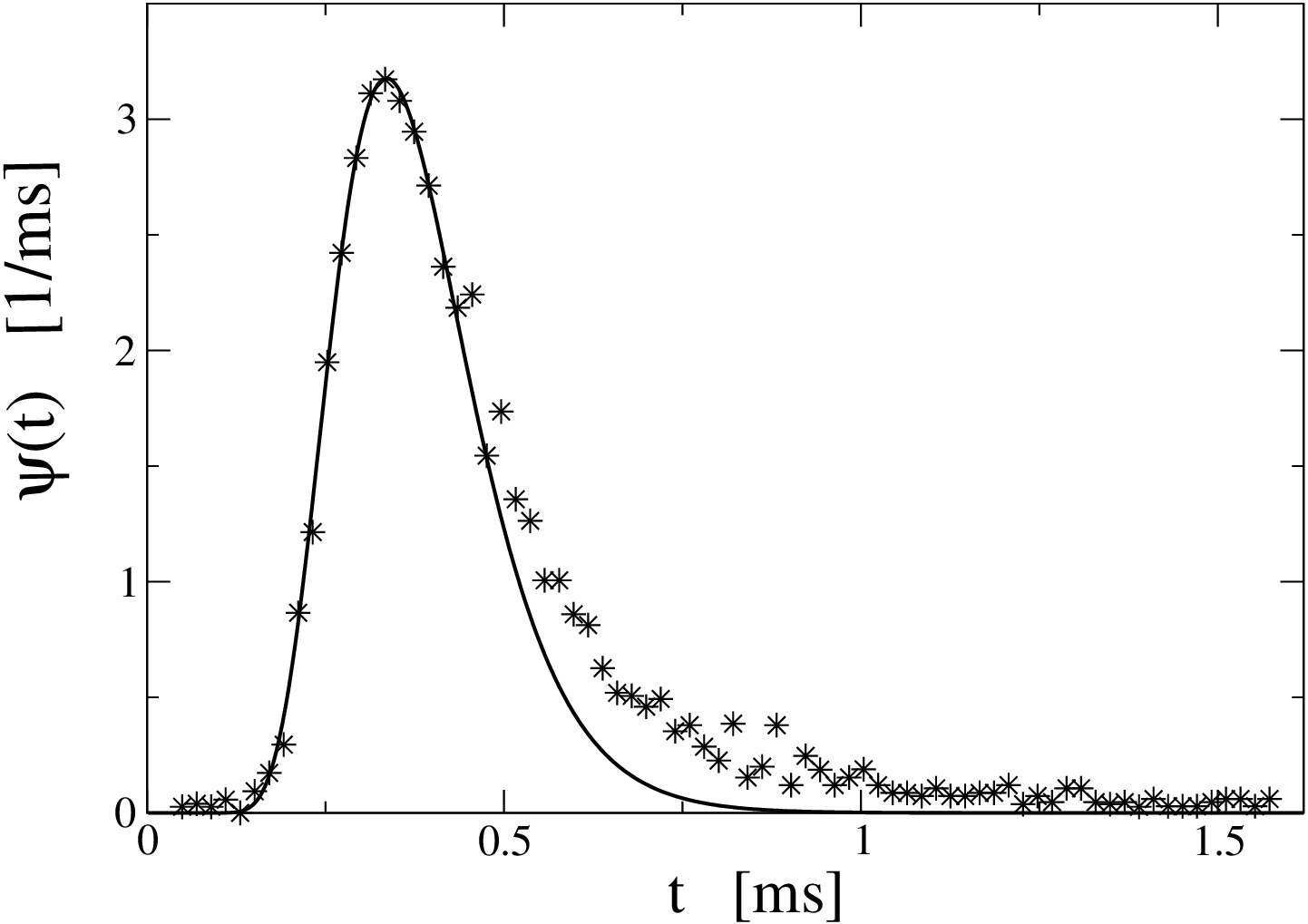} 	
      \caption{}
      \label{fig4}
   \end{center}
\end{figure}

\clearpage
\begin{figure}
   \begin{center}
   \epsfxsize=0.8\columnwidth
	\epsfbox{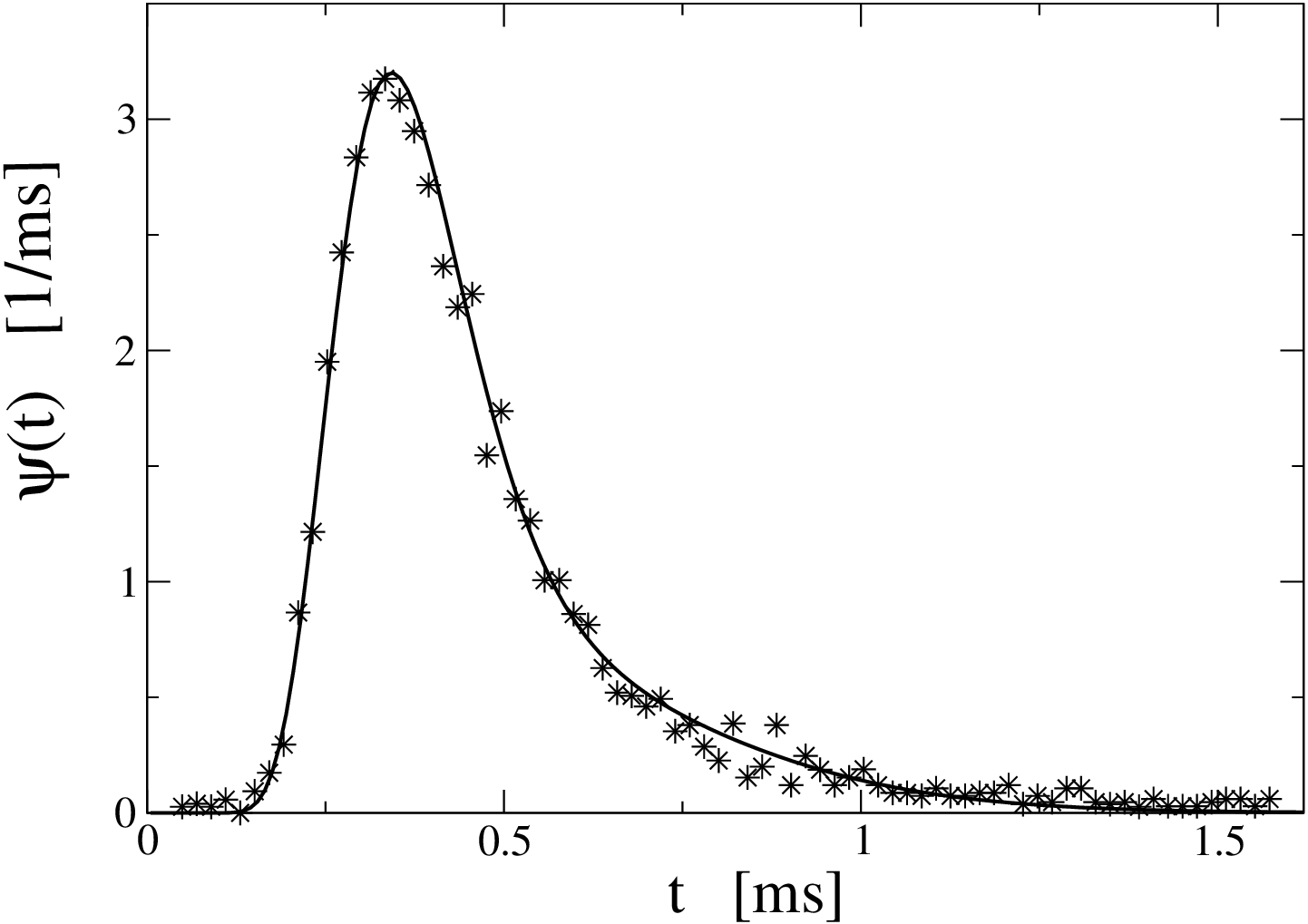} 	
      \caption{}
      \label{fig5}
   \end{center}
\end{figure}

\clearpage
\begin{figure}
   \begin{center}
   \epsfxsize=0.8\columnwidth
	\epsfbox{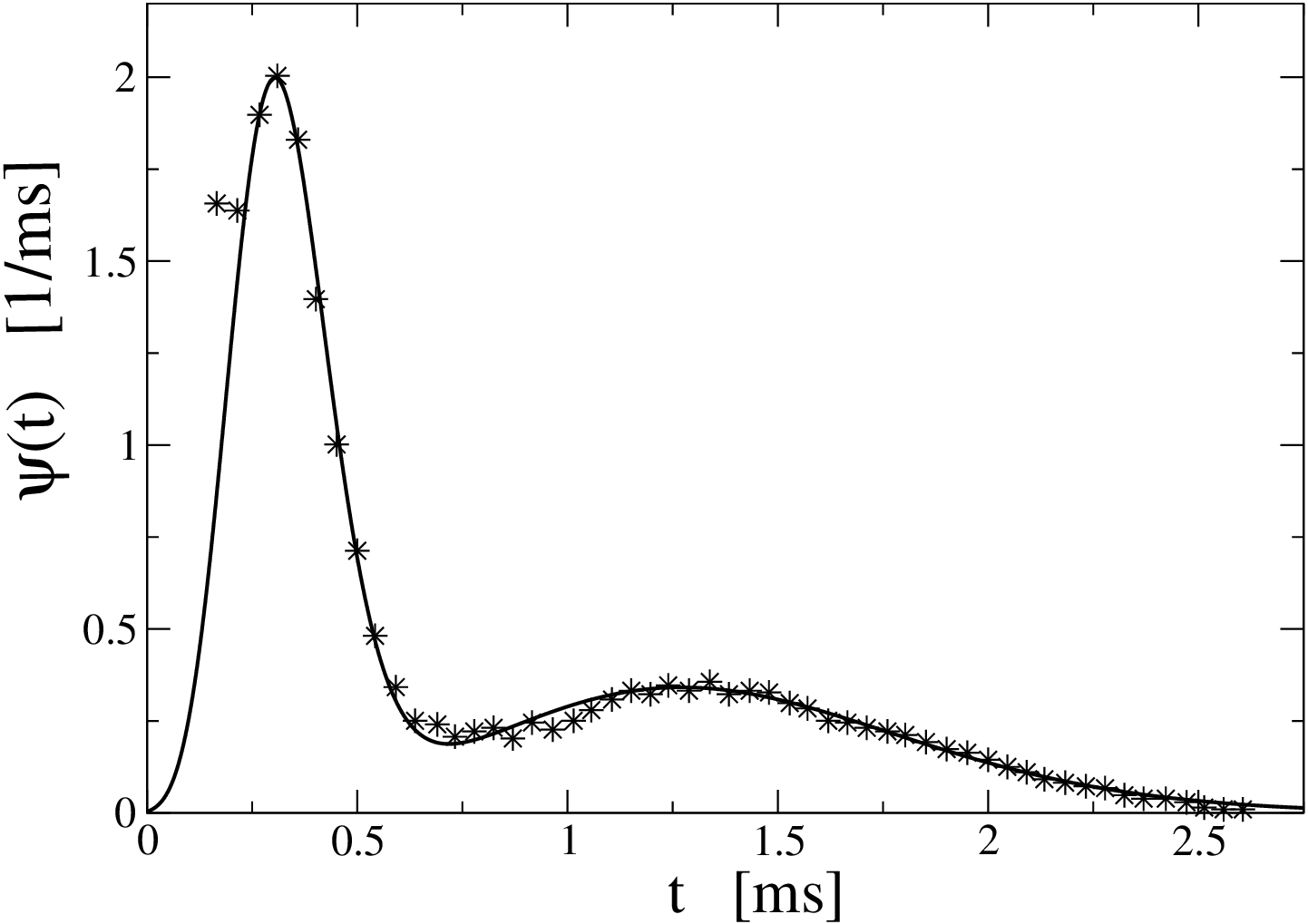} 	
      \caption{}
      \label{fig6}
   \end{center}
\end{figure}

\end{document}